# Enhanced Crystallization and Evaporation Retardation in Mixed Surfactant Systems at the Air-Water Interface: A Study on Chain Length Compatibility and Molecular Ratio

Kulsuma Begum[1,#], Abhijeet Das[1,2,#], Dinesh O. Shah[3,4], Sanjeev Kumar[1,4,*]

[1]Centre for Advanced Research, Department of Physics, Rajiv Gandhi University, Arunachal Pradesh 791112, India

[2]Department of Bioengineering, Indian Institute of Science, Bangalore – 560012, India

[3]Center for Surface Science and Engineering, Department of Chemical Engineering, University of Florida, Gainesville 32611, Florida

[4]Shah-Schulman Center for Surface Science and Nanotechnology, Dharmsingh Desai University, Nadiad, Gujarat

[#] Equal Contribution

* Corresponding Author: sanjeev.kumar@rgu.ac.in

[2] Current Affiliation - Department of Bioengineering, Indian Institute of Science, Bangalore – 560012, India

**Abstract:** Effects of chain length compatibility and molecular ratio on the two-dimensional crystallization of a binary mixed surfactant system with non-identical molecular size and its consequence on retardation to water evaporation are described via Langmuir Blodgett films. The mixed monolayers corresponding to 1:3 exhibit minimal area per molecule owing to identical chain length. The maximum crystallization was also observed at this ratio from BAM images at constant surface pressure. The prominent changes in the physical properties of the analyzed system, for the 1:3 molecular ratio, are attributed to the augmented stability mediated by the hexagonal closed packing and packing behavior in the mixed monolayer. The observation was validated from a random ball mixing model and simulation study. The maximum retardation to evaporation was also observed for the 1:3 molecular ratio and is attributed to augmented stability and spreading of the monolayers at the air-water interface.

# Introduction:

Manipulation of matter on a miniature scale has attracted the attention of scientists for the in-depth understanding of nanoscience and the fabrication of nanodevices and technologies. The self-assembly of macromolecules or particles into desired structures and interconnecting them is an obligatory step for the microfabrication of devices. Many industrial applications encounter the need for surfactants due to the presence of both polar and non-polar groups in a single molecule. In addition, a mixture of surfactants is highly desirable given their enhancement in performance as compared to a single surfactant system[1]. In this regard, two-dimensional crystallization (2D) of a concoction of surfactant molecules or particles is exceedingly necessary given the possibilities of generating a locally high concentration of surfactant constrained in 2D. A Langmuir monolayer acts as an excellent model for investigating the ordering or crystallization in two-dimensional layers deposited onto an ideally smooth substrate of water via systematic variation of surface pressure.

Earlier research reported that the stoichiometric association between the molecules in a mixed surfactant system can strikingly alter the system properties[1–6]. In particular, prominent variations in the properties of mixed surfactant systems were observed in the 1:3 molecular ratio due to the hexagonal close packing between the molecules[2,3,7–9]. Patist *et al*.[8] reported the minimum surface tension and rate of evaporation and the maximum surface viscosity, foam stability, and single film loop stability for the 1:3 molecular association of mixed cationic and anionic surfactant systems. In addition, maximal solubilization of water in microemulsion (oil-water interface) was observed for the 1:3 ratio of potassium oleate and hexanol[9]. Booij[4] showed a minimal penetration of oil molecules to the surfactant film at a 1:3 molecular ratio resulting in maximum extraction of oil from the emulsion.

A different imperative aspect to consider for interfacial monolayer films in mixed surfactant systems is their chain length compatibility i.e., the chain length of one surfactant should be well-matched with the chain length of another surfactant. It is for the cause of maximizing the molecular interaction in the lateral direction resulting in the stabilization of the interface. Since surface-active and additional hydrocarbon molecules are aligned at the interface, the interfacial properties are largely impacted by the matching and mismatching of chain lengths. In case of mismatch, there is a larger probability and freedom for the surplus hydrocarbon tails to disrupt the molecular packing via enhanced tail motion contributing to the propagation of disturbance, conformational disorders, etc. The understanding of this factor is crucial for tuning properties like foamability, micellar lifetime, surface viscosity, surface tension, bubble radius, contact angle, corrosion, oil recovery, microemulsion stability, etc.[10–12]. Previous investigations showed that in the case of matching between the adjacent hydrocarbon chain, there occurs an enhancement in the order, packing, and stability of hydrocarbon layers which further leads to maximization in the foamability, surface viscosity, and micellar lifetime. Consequently, due to the tight packing between the molecules, there follows a decrement in the bubble radius, surface tension, and contact angle[11]. Additionally, from the technological viewpoint probing the influence of chain length compatibility in surfactants will help to solve the problems regarding lubrication, oil recovery, corrosion, environmental remediation, and control of evaporation. Further, there is a high significance in understanding the retardation of water evaporation by monolayers of long-chain molecules both in the fundamental aspect of permeation of vapors as reproduced in the physical states of monolayers[13] and practical aspect for reducing the losses in natural reservoirs due to evaporation via fabrication of hydrophobic monolayers of long-chain alcohols and fatty acids taking water as substrate[14–19]. Nonetheless, only a few studies have been systematically conducted to investigate the effect of monolayer spreading on retardation to evaporation[14,20]. The aforementioned and possible correlation between chain length compatibility and molecular ratio with the two-dimensional crystallization of a mixed surfactant system can be investigated from a Langmuir-Blodgett (LB) system by incorporating Brewster Angle Microscopy (BAM) to relate the structural change to the corresponding change in compressibility derived from the simultaneously measured surface pressure-area ($\pi$-A) isotherm. Nonetheless, to the best of our knowledge, a systematic investigation on the effect of molecular ratios and hydrocarbon chain length on two-dimensional crystallization of a mixed surfactant system and its consequence on retardation to water evaporation is lacking in the literature.

Herein, we studied the 2D crystallization in a mixed surfactant system of Stearic acid ($C_{18}$) and Behnic alcohol ($C_{22}$) with varying molecular ratios by surface pressure-area isotherm and Brewster angle microscopy analysis. The maximum crystallization was observed in 1:3 molecular ratio of $C_{18}$ acid and $C_{22}$ alcohol at a surface pressure of 35mN/m. Consequently, the system has been applied to probe its effect on the retardation of water. In addition, atomic force microscopy (AFM), random ball mixing model, and simulation have been applied to verify the augmented ordering of mixed surfactant molecules at a 1:3 ratio. It is noteworthy that owing to the four-carbon difference in chain length between $C_{18}$ and $C_{22}$ resulting in 0.6 *nm* indentation or hole in the mixed monolayer surfaces, we utilized the AFM technique owing to its sensitiveness down to the length of 0.1 *nm* in the *z-direction*.

## Experimental Details:

Solutions of two surfactants; Stearic acid ($C_{18}$) and Behnic alcohol ($C_{22}$) are prepared in ultra-clean glassware with a concentration of 1 mg in 1 ml in a mixture of Chloroform, Methanol, and Hexane (1:1:3). The solutions were mixed in different molecular ratios (1:0 1:1, 1:2, 0:1, 1:3, 2:1 and 3:1) and spread onto the water substrate in Langmuir-Blodgett trough (KVS Nima). The films were compressed very slowly after 10 minutes and a Brewster angle microscope (KVS Nima) was employed to capture the monolayers' different surface pressure after holding for 2 minutes at the air-water interface. Atomic force microscopy (NT-MDT Ntegra Prima) was utilized to image the molecular arrangement in the film surface by transferring it to mica and the silica absorption method was employed to probe the effect of maximum crystallization at constant pressure on retardation of water evaporation. All the measurements were carried out at room temperature.

## Results and Discussion:

### Π-A Isotherm:

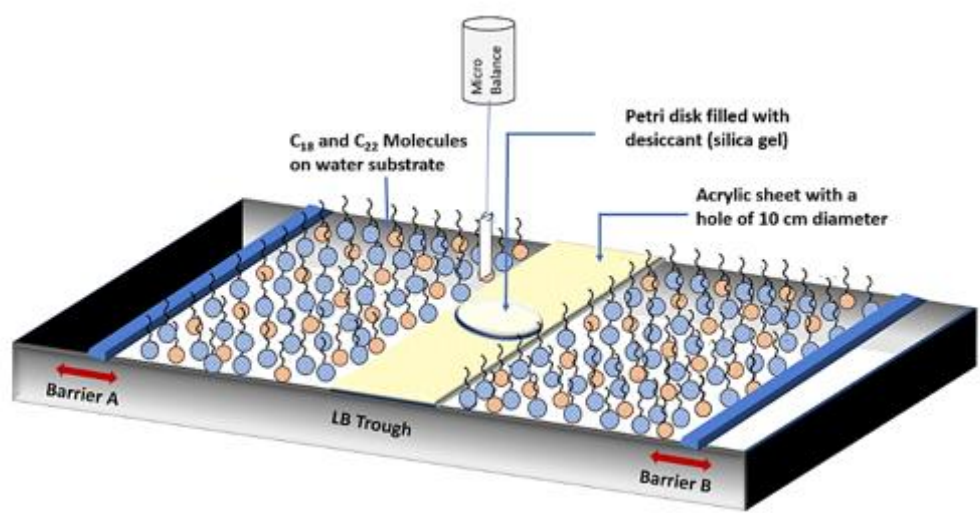

**Fig. 1** Pictorial representation of the experimental setup

The isotherm characteristics for different mixing ratios of Stearic acid $C_{18}$ and Behnic alcohol $C_{22}$ molecules were studied by spreading the solutions on the water surface in the LB trough. The experimental setup is displayed in Fig. 1. The isotherms exhibited a low-pressure 2D liquid-expended phase followed by a liquid-condensed or solid region. The isotherm collapses slightly below 50 *mN/m* for $C_{18}$ (1:0) and falls at 30 *mN/m* for $C_{22}$ (0:1) as shown in Fig. 2. The higher collapse pressure in 1:0 is presumably due to the refusal of polar group area to squeeze the water between the molecules or around the polar groups resulting in some molecules getting squeezed out of the monolayer and falling on the top of adjacent one thus forming bi-, tri-layers or micelle type aggregation. On the contrary, the strong cohesion/adhesion between the hydrocarbon chains lifts the monolayer resulting in a lower collapse pressure for a 0:1 molecular ratio. However, no uniformity has been observed in the collapse surface pressure of mixed monolayers suggesting its dependence upon the molecular packing in mixed monolayers. This can be elucidated from the phase rule according to which the two components in a mixed monolayer are miscible at the interface due to the variation of surface pressure with composition. However, the appearance of the phase transition points in Fig. 2 similar to that of pure compounds in the mixed monolayers suggests the existence of pure $C_{18}$ (or $C_{22}$) molecules in the mixed monolayer i.e., the components in the mixed monolayers are not completely miscible at the air-water interface.

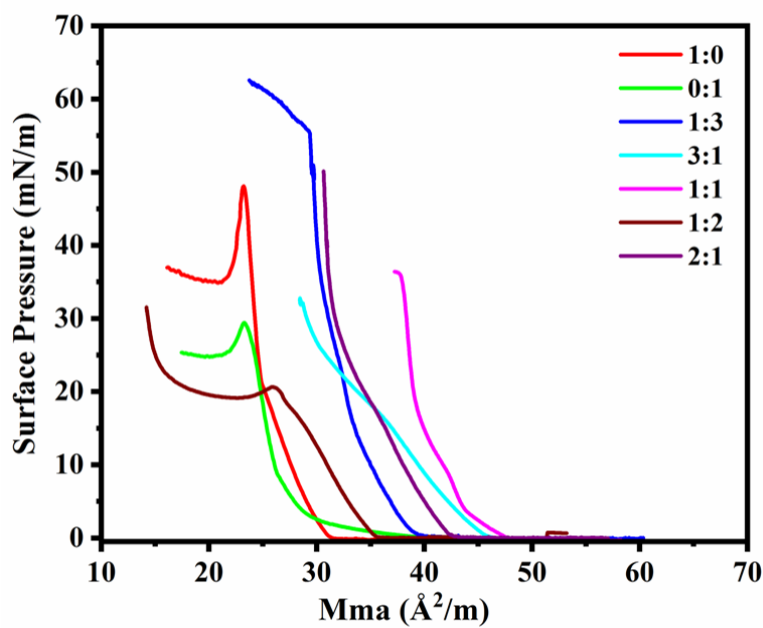

**Fig. 2** π-A isotherms of mixed surfactants at different molecular ratios

For the 1:1 molecular ratio, for a homogeneously mixed solution, there is a presence of both hydroxyl and carboxyl groups in 50-50%. Due to the possibility that one alcohol group might not be immediately next to another alcohol group, there is a disturbance due to the thermal motion of four carbon tails which further will propagate down resulting in an increase in the area per molecule to around 40 Å²/molecule. Also, the collapse of the monolayer at surface pressure between those of 1:0 and 0:1 might be because the mixture partly consists of both alcohol and carboxyl groups and there exists an asymmetry between the adjacent molecules due to a large number of four-carbon chains. In addition, 67 molecules of long-chain alcohol and

33 molecules of fatty acids are present for the 2:1 mixing ratio. Due to this, it is easier for the alcohol group to squeeze out the monolayer to air and as a consequence, the polar groups will come out due to their molecular motion resulting in a quicker collapse. Interestingly, in this case, a decrement in available surface area per molecule is observed compared to the 1:1 molecular ratio. This observance can be due to the presence of maximum disorder resulting in the hydration of groups and consequently a larger area. For a molecular ratio of 1:2, the collapse pressure and surface area are observed to be around $15\ mN/m$ and 25 Å²/molecule, respectively. Disorder in the vertical chain mediated by the presence of more hydroxyl groups as compared to carboxyl results in easy lift-off of monolayer resulting in a faster collapse and minimal surface area.

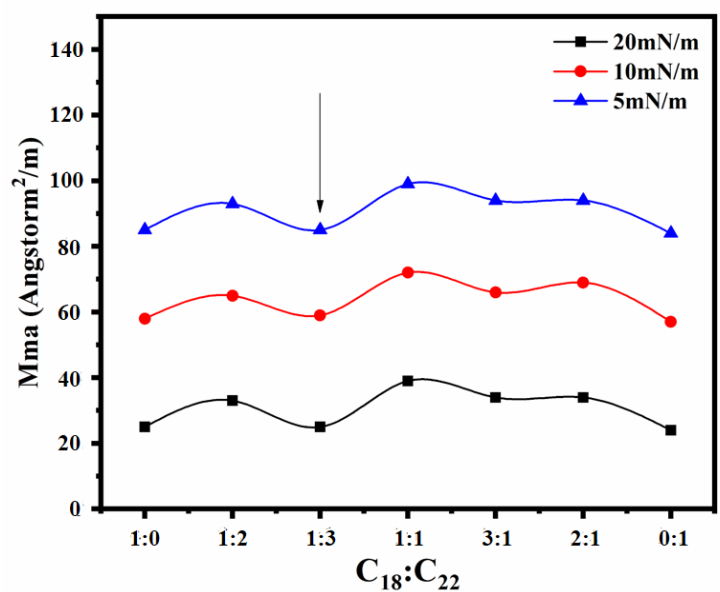

**Fig. 3** Variation of accessible molecular area with mixing ratio at constant surface pressure for the mixed surfactant system

Besides, the inflection (breaking) point reaches its largest value of 55 *mN/m* for the molecular ratio of 1:3 indicating maximum 2D crystallization or ordered arrangement of hydrocarbon chain for this mixing ratio. Additionally, this mixing ratio has lesser accessible area as compared to other ratios at constant pressure as depicted in Fig. 3 which indicates an enhancement in molecular order mediated by the identical length of molecular chains resulting in their highly ordered arrangement and consequently, compensating the disturbance produced from the terminal portion of the molecules. In addition, this also suggests the presence of maximum surface viscosity and minimum surface tension for a system of mixed surfactants at 1:3 ratio of molecules. Finally, larger molecular area at molecular ratio of 3:1 is attributed to the disturbance from the motion of the excess carbon tails resulting in a faster collapse of the monolayer[2,8].

Also, Fig. 2 demonstrates the existence of different compression profiles and phase formation around the ratio of different chain length molecules. Specifically, at a molecular ratio of 1:3 liquid-expanded phase is detected, which has transformed into a condensed phase at low pressures. A plateau region of co-existing expanded and condensed phase follows, consistent with the first-order nature of the expanded-condensed transition[21,22]. The situation is different at ratios other than 1:3. This shows the tightly closed-packed condensed phase destabilizes the crystallization. The trend of short relaxations at large areas and longer relaxations at smaller areas, visible here, was observed throughout the study. This trend has been explained[23]. At large areas, the film is made up of islands of molecules in co-existence with a gaseous phase and as the area gets reduced and the islands are squeezed together, relaxation effects are dominated by local shape changes necessitated by the points where islands come into contact. These are necessarily more rapid than macroscopic, non-local grain boundary relaxations that become requisite as the molecular density saturates. Significant compression beyond saturation requires some removal of molecules from the surface.

**BAM Analysis**: The Brewster Angle Microscopy (BAM) technique was used to capture the real-time crystallization information of the monolayer of Behnic alcohol and mixed monolayers of $C_{18}$ and $C_{22}$ at 1:3 and 3:1 molecular ratio, respectively after holding for 2 minutes from compression as displayed in Fig. 4. In the pristine $C_{22}$, no or little light is reflected from the air-water interface due to large area occupied by molecules and thus dark images was observed thus, identified to be the gas phase. In addition, bright zones of low contrast and brightness appear upon compression of monolayers indicating an increment in molecular density and resulting in the transition from gaseous to liquid phase. This observation supports the quicker collapse of monolayer film as observed in the $\pi - A$ isotherm analysis. For the mixed monolayers at 1:3 ratio as compared to 3:1, dots with enhanced brightness and uniform phases are seen at 35 mN/m suggesting an augmented crystallization and formation of condensed or solid phase with compact domains. The transition from expanded to condensed phase occurs with increasing pressure due to the increase in stability of the condensed phase as a result of the Brownian motion of the molecules[24].

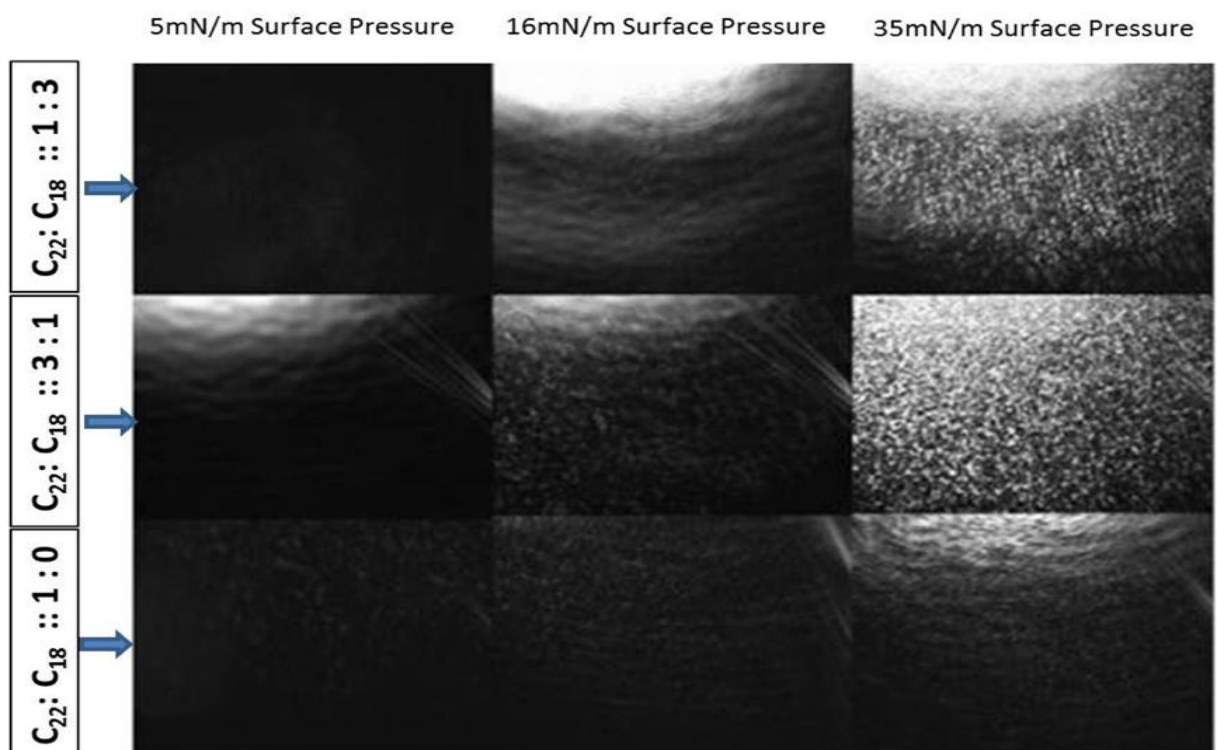

**Fig. 4** BAM images corresponding to pristine $C_{22}$ and mixed monolayers at 1:3 and 3:1 molecular ratio, respectively with varying surface pressure

**Two-dimensional closed packing:** A fundamental factor is liable regarding the striking changes in the physical properties of mixed surfactant system at 1:3 molecular ratio as reported in earlier works[2,8].

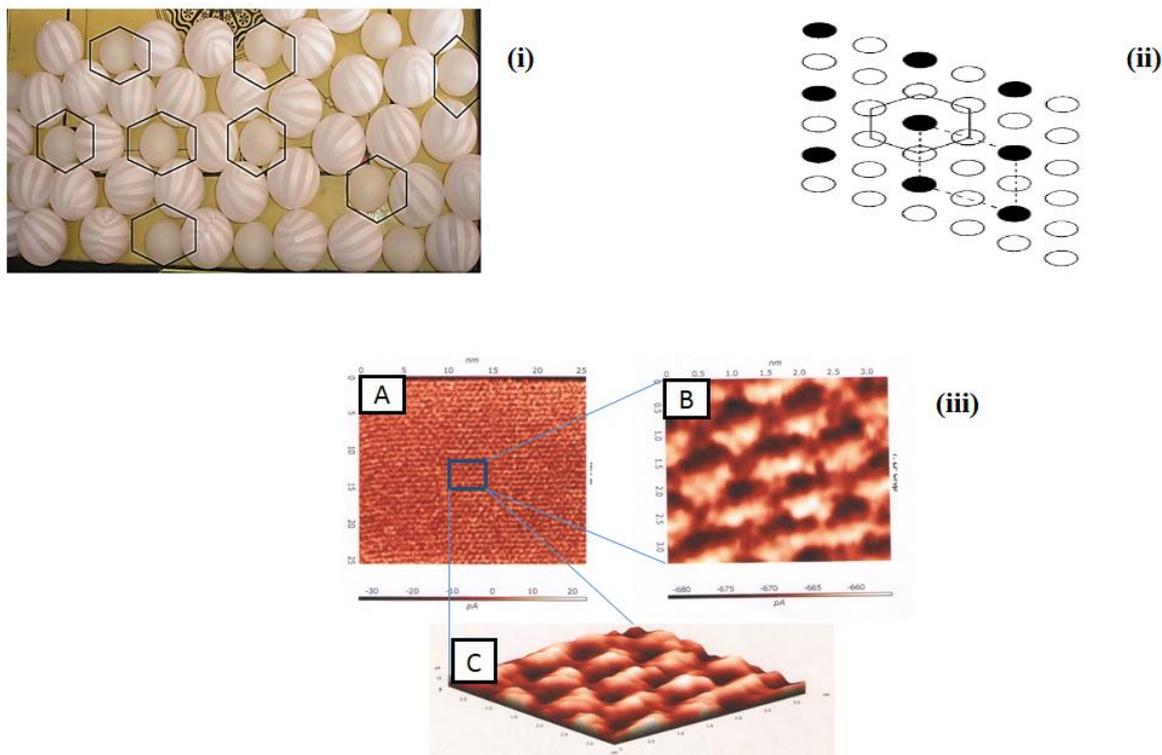

**Fig. 5 (i)** Formation of hcp structure in random mixing of two sets of balls with different radii, **(ii)** Proposed arrangement of different sized molecules in a mixed surfactant system[2], and **(iii)** (A-B) 2D and (C) 3D AFM pictograph of mixed monolayer at 1:3 molecular ratio

In order to investigate the association between the binary system of molecules with non-identical area and possible patterns in random packing, a ball experiment was performed with varying ratios from two sets of balls with different sizes or radii on a charm board. Fig. 5(i) displays the formation of hexagonal closed packed (hcp) structures in random mixing of a binary system of balls or spheres with different radii where, the maximum number of hexagonal patterns are observed in case of a 1:3 ratio of small to big balls, respectively. The observation supports the proposed possible arrangements of molecules displayed in Fig. 5(ii)[2]. Due to the four-carbon difference in chain length between $C_{18}$ and $C_{22}$ resulting in 0.6 *nm* indentation or hole in the mixed monolayers, we utilized AFM owing to its sensitivity to the length of 0.1 *nm* in the z-direction. In this regard, the AFM image of the mixed monolayer at a 1:3 molecular ratio as displayed in Fig. 5(iii) exhibits the presence of hexagonal packing consequently, validates the significant dependency of the degree of association in molecules on the molecular area. From the aforementioned, it can be inferred that chain length compatibility in mixed surfactant systems is a prominent factor in the enhanced ordering of molecules. Furthermore, two different sizes of the molecule can be arranged in monolayers such that molecules of one type occupy the centers and those of the other type occupy the corners to form a hexagonal pattern[2]. This nature of packing would yield the minimal area per molecule and thus, would not make the 1:3 and 3:1 ratio interchangeable between the types of molecules. So, it is suggested that although hexagonal packing is possible in monolayers of molecules with identical areas or sizes for mixed systems consisting of molecules with non-identical areas, maximum packing is possible only in case of a 1:3 association of molecules.

To further validate the observation of augmented close packing in a mixed surfactant system with a 1:3 molecular ratio, we studied the packing efficiency and the increment in packing density for different molecular ratios by considering the system to be a close packing of mixed spheres (molecules)[25]. The efficiency in packing for different molecular ratios is calculated from the radius ratio using, $\frac{r}{R} = \frac{\sqrt{2}-1}{\sqrt{n}}$ while the increment in density is calculated from $n\left(\frac{r}{R}\right)^3$; $r$ and $R$ being the radii of smaller and larger molecules (spheres), respectively and $n$ is the number of smaller spheres.

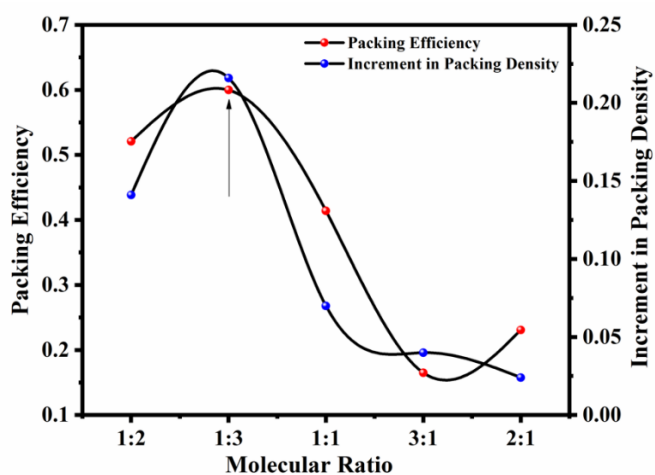

**Fig. 6** Variation of packing efficiency and incremental packing density with molecular ratio

The double Y-axes plot between molecular ratios versus packing efficiency and increment in packing density is shown in Fig. 6. The highest and lowest efficiency in packing and increment of packing density is observed for the 1:3 and 3:1 molecular ratio, respectively. The packing efficiency at a 1:3 molecular ratio resembles significantly that of the hexagonal closed-packed arrangement. The highest packing efficiency (~0.6) is due to the occupancy of the corners by the larger spheres and the interstitial site or centre by the smaller sphere while for the lowest efficiency could be due to the occupancy of larger sphere at the face centre. The dependency of molecular association on the arrangements of surfactant molecules was suggested by Shah

*et. al.*[2]. In this regard, the formation of hcp structures for the molecular ratio of 1:3 only is due to the non-identical surface area of the molecules whereas, the enhanced structural ordering is due to the enhanced intermolecular interaction from a reduced spacing in the intermolecular distance. This characteristic is also attributed to the lowest packing efficiency of the 3:1 molecular ratio as the larger molecule occupies the greater fraction[2].

We also simulate the random mixing of a set of spheres for individual ratios utilizing the libraries in Python to authenticate the observations. For the simulation, we generated the coordination of the spheres using the normal distribution between -50 and 50 in each dimension. Next, we checked for possible overlaps between the spheres using the distance formula and accordingly, adjusted the position to remove them. Finally, we computed the packing density and packing fraction, respectively for each of the mixing ratios (Table 1) and plotted the mixing in three dimensions, as displayed in Fig. 7. The simulation code corresponding to the 1:3 molecular ratio is provided (S1 see supplementary material). The largest packing density and packing fraction corresponding to 1:3 is evident from Table 1 consequently, validating the observed maximum two-dimensional crystallization in a mixed surfactant system facilitated by identical molecular chain length. In addition, the obtained results qualitatively complement the observation from Fig. 6. The obtained results confirm the significance of chain-length compatibility and molecular ratio of 1:3 in the two-dimensional crystallization of systems consisting of mixed surfactants.

**Table 1.** Variation of packing density and packing fraction in random mixing of spheres corresponding to the mixing ratio

| Mixing Ratio | Packing Density | Packing Fraction |
|---|---|---|
| 1:1 | 0.0005 | 0.9 |
| 2:1 | 0.0005 | 0.8 |
| 1:2 | 0.0009 | 0.94 |
| 3:1 | 0.0005 | 0.73 |
| 1:3 | 0.001 | 0.96 |

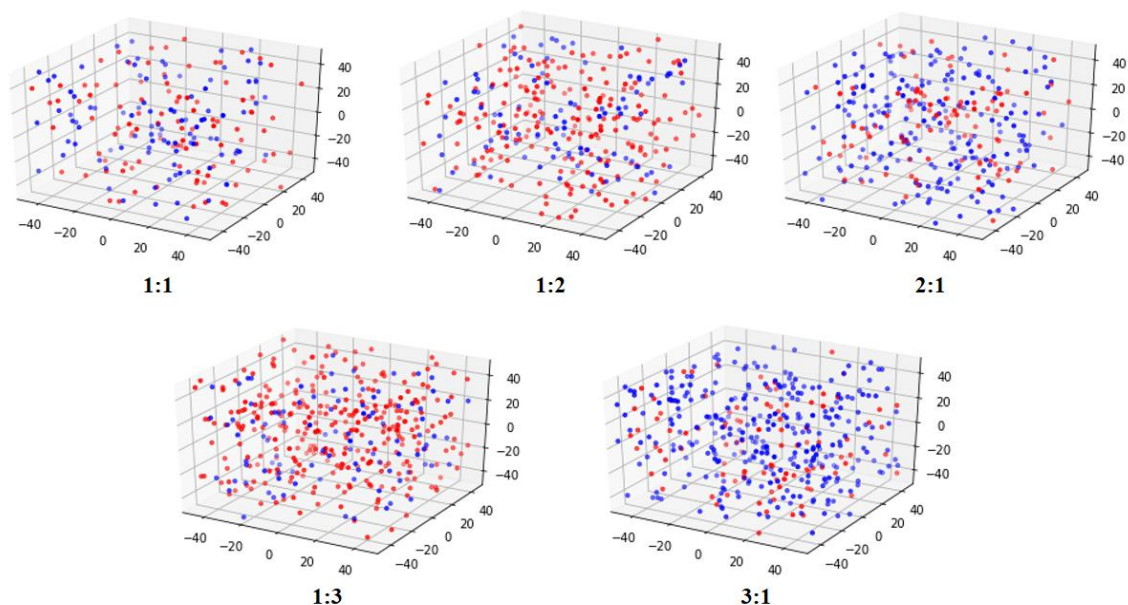

**Fig. 7** Simulated random mixing of small (blue) and large (red) spheres in different ratio

Nonetheless, we employed a simplified three-dimensional ball-mixing model to illustrate the packing efficiency at different mixing ratios conceptually. While the actual system is two-dimensional, this analogy effectively demonstrates that a specific ratio of differently sized components can lead to enhanced packing density compared to random mixing or ratios with a large excess of one component.

**Retardation of Evaporation Analysis**: The influence of crystallization facilitated by the molecular ratio on the rate of evaporation was investigated from the LB system. The information regarding the variation in the rate of evaporation at a surface pressure of 0, 15, 30, and 40 *mN/m*, respectively is provided in Fig. S1. The trend in retardation of evaporation for different molecular ratios at a constant surface pressure of 30 *mN/m* and varying surface pressure at 1:3 molecular ratio is shown in Fig. 8(a) and 8(b), respectively. The augmented retardation in water evaporation at 1:3 is evident from Fig. 8 and found to be ~ 45%. According to Simko and Dressler[26], longer chain lengths should give higher efficiencies however, no particular trend is observed for the retardation indicating that the molecular arrangement has a significant role in this process, i.e. the molecular arrangement may allow molecules of one carbon to fit between molecules of other carbon thus preventing the water molecules to escape.

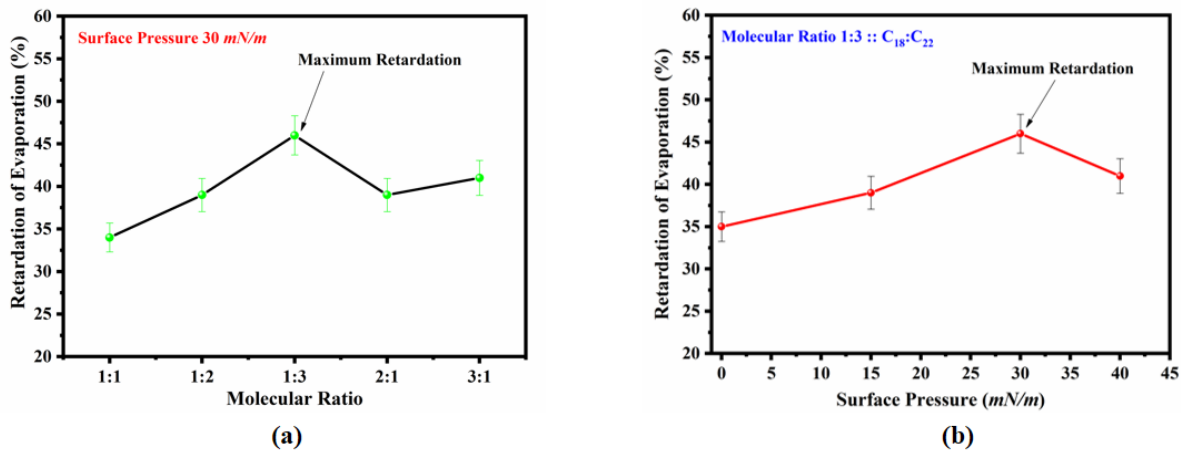

**Fig. 8** Variation in retardation to water evaporation with **(a)** molecular ratio at constant surface pressure and **(b)** surface pressure corresponding to 1:3 molecular mixing ratio

Additionally, the intrinsic resistance of the monolayers and its spreading decides the degree of effectiveness of the monolayer in the retardation process[26]. In this context, BAM pictures corresponding to 3:1, as shown in Fig. 4 confirm that the monolayers have the highest spreading thus, increasing the retardation of evaporation. Furthermore, it is also observed that carbon numbers above 18 tend to break under high compression[26]. This implies that the molecular arrangement at 1:3 may possess more fluidity consequently, have a less brittle nature and hence more stability. Moreover, the maximum retardation to evaporation also indicates minimum surface tension and maximum viscosity[8] following the results from π-A isotherm analysis.

## Conclusion:

The present study described the effect of chain length compatibility and molecular ratio on the 2D crystallization in mixed monolayers via Langmuir-Blodgett films. The π-A isotherm analysis revealed the minimal accessible area per molecule and maximum surface pressure for the mixed monolayer of $C_{18}$ and $C_{22}$ at a molecular ratio of 1:3 facilitated by the identical hydrocarbon chain length. The real-time captured BAM images presented the maximum crystallization for this ratio at a surface pressure of 35 *mN/m*. Atomic force microscopy study exposes the striking changes in the physical properties of the analyzed mixed surfactant system from the augmentation in stability of mixed monolayers mediated by the ordering of molecules or hexagonal closed packing at a 1:3 mixing ratio. The observation is qualitatively supported by the random mixing of balls and simulation-based study. The retardation to water evaporation investigation revealed the significance of molecular arrangement in this process. The maximum retardation to evaporation was observed for the 1:3 molecular ratio and is attributed to augmented stability and spreading of the monolayers at the air-water interface.

**Conflict of Interest:** The authors declare that they have no known competing financial interests or personal relationships that could have appeared to influence the work reported in this paper.

**Data Availability Statement:** The data supporting the findings of this study are available from the corresponding author upon reasonable request.

# Supplementary Information

# Enhanced Crystallization and Evaporation Retardation in Mixed Surfactant Systems at the Air-Water Interface: A Study on Chain Length Compatibility and Molecular Ratio

Kulsuma Begum[1,#], Abhijeet Das[1,2,#], Dinesh O. Shah[3,4], Sanjeev Kumar[1,4,*]

[1]Centre for Advanced Research, Department of Physics, Rajiv Gandhi University, Arunachal Pradesh 791112, India

[2]Department of Bioengineering, Indian Institute of Science, Bangalore – 560012, India

[3]Center for Surface Science and Engineering, Department of Chemical Engineering, University of Florida, Gainesville 32611, Florida

[4]Shah-Schulman Center for Surface Science and Nanotechnology, Dharmsingh Desai University, Nadiad, Gujarat

[#] Equal Contribution

[*] Corresponding Author: sanjeev.kumar@rgu.ac.in

[2] Current Affiliation - Department of Bioengineering, Indian Institute of Science, Bangalore – 560012, India

```python
import numpy as np

import matplotlib.pyplot as plt

from mpl_toolkits.mplot3d import Axes3D

# Define the radii and number of spheres

r1 = 1.0

r2 = 0.5

n1 = 300

n2 = 100

# Generate the coordinates of the spheres randomly using a uniform distribution

coords = np.random.uniform(-50, 50, (n1 + n2, 3))

# Check for any overlaps between the spheres and adjust the position of the spheres if necessary

for i in range(n1 + n2):

    for j in range(i + 1, n1 + n2):

        dist = np.linalg.norm(coords[i] - coords[j])

        if dist < (r1 + r2):

            overlap = (r1 + r2) - dist

            adjust = overlap / 2

            direction = (coords[j] - coords[i]) / dist

            coords[i] -= adjust * direction

            coords[j] += adjust * direction

# Calculate the packing density and packing fraction

v1 = (4/3) * np.pi * r1**3

v2 = (4/3) * np.pi * r2**3

v_total = n1 * v1 + n2 * v2
```

```python
density = v_total / (100**3)
fraction = n1 * v1 / v_total
# Plot the result in 3D
fig = plt.figure()
ax = fig.add_subplot(111, projection='3d')
ax.scatter(coords[:n1, 0], coords[:n1, 1], coords[:n1, 2], c='r', s=10)
ax.scatter(coords[n1:, 0], coords[n1:, 1], coords[n1:, 2], c='b', s=10)
ax.set_xlim(-50, 50)
ax.set_ylim(-50, 50)
ax.set_zlim(-50, 50)
plt.show()
print("Packing density:", density)
print("Packing fraction:", fraction)
```

**S1.** Python program to simulate the random 1:3 mixing of non-overlapping and non-identical spheres and compute the packing density and packing fraction of the mixing, respectively

## Measurement of the rate of water evaporation at different surface pressure

The measurements are made in the Langmuir trough. The experimental setup is shown in Fig. 1 of the article. The desiccant container is a petri dish, on the open side, covered with a piece of porous cloth that retains the desiccant but is permeable to water vapor. An acrylic sheet with a hole of 10 *cm* diameter is placed on the trough which rests on the edge of the trough which holds the petri dish parallel to and about 2 *mm* from the water surface.

To determine the rate of evaporation, the petri dish, filled with silica gel (desiccant) is weighed and put into the position above the water surface for a definite length of time. Thereafter, the petri dish is removed and reweighed. The procedure is repeated several times for the mixed monolayers of Stearic acid ($C_{18}$) and Behnic alcohol ($C_{22}$) with different molecular ratios (1:1, 1:2, 1:3, 2:1, and 3:1) at surface pressures 0, 15 *mN/m*, 30 *mN/m* and 40 *mN/m* as shown in Fig. S1. The surface pressure is maintained by the movable barriers A and B using LBXD software. All the observations have been recorded at room temperature.

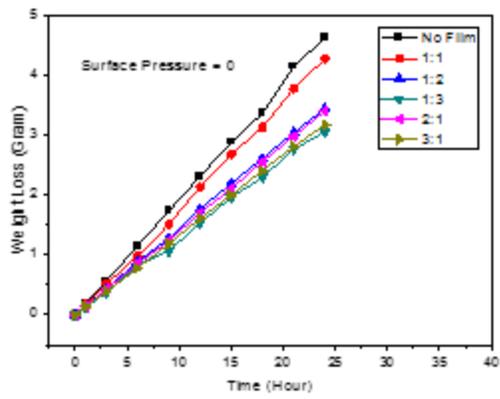 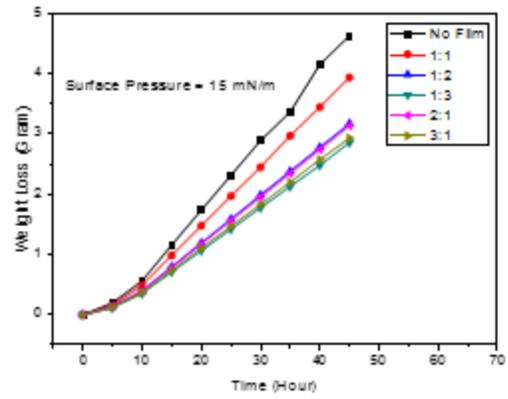
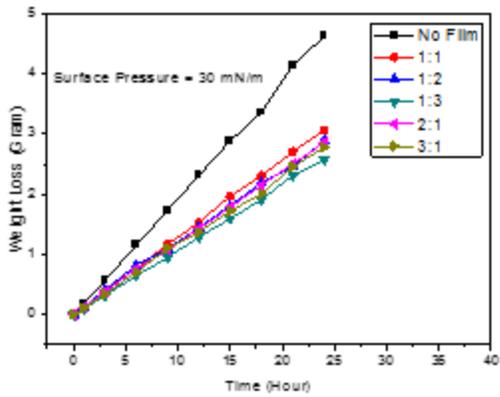 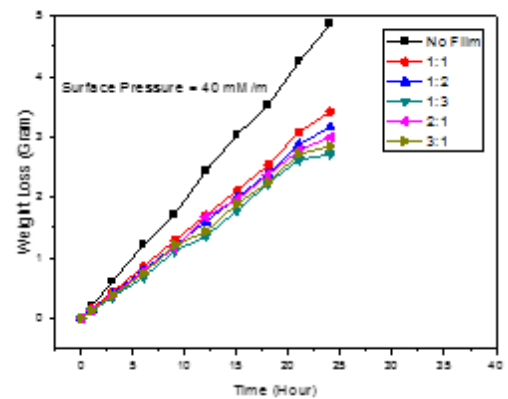

**Fig. S1** Rate of evaporation at surface pressure 0, 15 *mN/m*, 30 *mN/m* and 40 *mN/m* for the analyzed molecular ratios